\documentclass[twocolumn,showpacs,prl]{revtex4}
\usepackage{graphicx} 

\begin{document}
{\bf Comment on ``Microscopic theory of network glasses''}
\par
In a recent Letter Hall and Wolynes\cite{wolynes} (HW) ask whether 
a microscopic
theory of network glasses can be developed starting from a model of
dense spherical fluids.  To do so they constrain the number of nearest
neighbors and count their central force interactions separately.  They
obtain the dynamical transition temperature $T_A$ (below which the system
is nonergodic and the motion is landscape-determined), and the entropy
crisis (Kauzman) temperature $T_K$ as functions of $n_b$, the average number
of nearest neighbor bonds/atom.  A Lindemann melting criterion on the 
amplitude of nearest neighbor vibrations defines the glass transition 
temperature $T_G$. The model shows that $T_A/T_G$ and $T_K/T_G$ monotonically 
increase and decrease, respectively, as functions of $n_b$.  For $n_b =
2.4$, $T_A/T_G = 100$, an unreasonable result.  Mode coupling theory (MCT) 
defines a critical nonergodicity temperature $T_c$ beyond which a radical 
change in the long time limit of e.g. the density-density correlation function occurs.  $T_c$ has been plausibly estimated for vitreous silica in 
molecular dynamics simulations\cite{Kob1} as $T_c/T_G = 2$. Even this
temperature is presently outside the reach of experimental 
investigations\cite{Kob1}.
\par
How do these results  compare with experiments ? There are standard
procedures for extrapolating specific heats to obtain $T_K$, but
identifying the onset of nonergodicity at $T_A$ (or $T_c$) is much
more difficult. Chalcogenide glasses
are ideal test systems because one can synthesize them over
a wide range of $n_b$ by chemical  alloying group IV  
additives in Se base glass. Fortunately, glass transitions of these 
systems have  been
recently reinvestigated \cite{r2} using  modulated DSC  (MDSC),  a new
method   which permits separating the  usual   DSC heat flow endotherm
$\dot H_T$ into a reversing part $\dot H_{rev}$  which is ergodic (and
which follows  the  modulated T-profile)  from the  non-reversing part
$\dot H_{nr}$ which  is non-ergodic (arising from underlying
temperature dependent activated
processes) as illustrated in Fig.1a. 
MDSC permits to establish  this
temperature  $T_A$ at  which  dynamics   become  landscape dominated
(i.e. in MDSC language dominated by a T-dependent $\dot H_{nr}$)
in contrast to the
linear response regime  (i.e. at high  temperatures when the  heat flow is
$\dot  H_{rev}$ dominated with a constant activation energy).  
\par
In binary $Ge_xSe_{1-x}$ glasses, observed variations in $T_A/T_G$ and
$T_K/T_G$  as a function of $n_b = 2 + 2x$ are compared to HW
predictions in Fig.1b. The $T_K/T_G (n_b)$ results were obtained from
a Vogel-Fulcher analysis of viscosity measurements\cite{r4}. And one finds
the observed and predicted variations in $T_K/T_G (n_b)$ ratio to be
in reasonable accord with each other showing a 
general reduction starting from a value of about $0.9$ at $n_b = 2$ to a
value of $0.8$ near $n_b = 2.7$. Note, however, that the broad global
minimum in the observed $T_K/T_G$ ratio near $n_b = 2.4$ in not
reproduced by the HW approach. More  serious is the fact that observed
$T_A/T_G$ values are (i) {\em two orders of magnitude lower than the HW
predictions} and (ii) {\em show a global minimum 
near  $n_b =  2.4$ that is in sharp contrast to the almost linear increase
(Fig. 1b) predicted  by HW}. The global minimum in $T_K/T_G$ and $T_A/T_G$ ratios
are features related  to self-organization of glasses that are missing in the 
HW approach. Clearly 
features such as inclusion of {\em local structures}\cite{r2,r5}, 
{\em structural
self-organization}\cite{r5,r7} and {\em non-central foerces
(bond-bending)} are missing in the theory. 
\begin{figure}
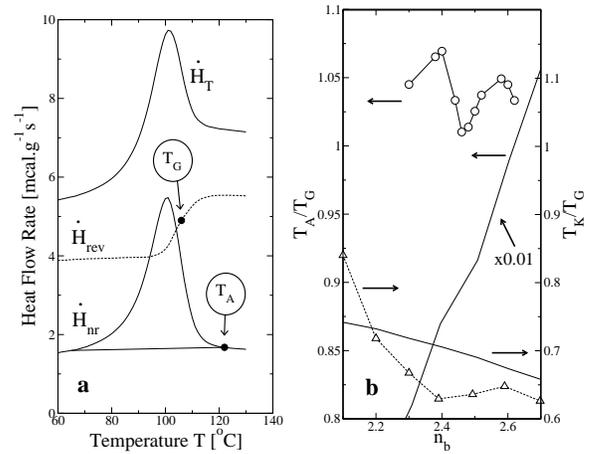

\resizebox{3.6cm}{!}{\includegraphics{as20bis.eps}}
\resizebox{3.9cm}{!}{\includegraphics{tatg2.eps}}

\caption{a: MDSC scan of $As_{20}Se_{80}$ bulk glass; $T_G$ is defined as
inflexion point of $\dot H_{rev}$ while $T_A$ is the end point of the
$\dot H_{nr}$ endotherm. b: $T_A/T_G$ ($\circ$) and $T_K/T_G$ ($\triangle$) for
Ge-Se glasses. Lines without
symbols are HW predictions [1]}
\end{figure}
Non-bending forces constitute $(4n_b-6)/(5n_b-6)$ of the
global number of network constraints (e.g. $0.6$ at the ideal $n_b=2.4$)
indicating that the directional non central angular forces have to be 
included in a successful theory of network glasses.
\par
M. Micoulaut$^1$ and P. Boolchand$^2$\par
\noindent
$^1$ LPTL, Universit\'e Paris VI, 4, place Jussieu 75252 Paris Cedex 05, France\par
\noindent
$^2$ Department of ECECS, University of Cincinnati, Cincinnati, Ohio 45221-0030

\end{document}